\documentclass[12pt]{article}
 \usepackage{amsfonts}
 \usepackage{amssymb,amsmath}
 \usepackage{graphicx} \usepackage{epstopdf}

 \newcommand{\crlb}[1]{\label{#1}\\[2pt]}
 
 \newcommand{\crld}[1]{\label{#1}}
 \newcommand{\eela}[1]{\quad\hbox{\scriptsize{#1}}\label{#1}\end{eqnarray}}
 \newcommand{\eelb}[1]{\label{#1}\end{eqnarray}}
 
 \newcommand{\newsecb}[2]{\section{#1}\label{#2}\setcounter{equation}{0}}

 \newcommand{\nolabels} {\def\eel{\eelb}\def\eeql{\eeqlb}  \def\crl{\crlb} 
 \def\newsecl{\newsecb}\def\bibiteml{\bibitem} \def\citel{\cite}\def\labell{\crld}}
\newcommand{\eeqla}[1]{\quad\hbox{\scriptsize{#1}}\label{#1}\end{aligned}\end{equation}}
\newcommand{\eeqlb}[1]{\label{#1}\end{aligned}\end{equation}}

\newcommand\publishversion  {\nolabels\setlength{\textheight}{8.38in}\setlength
    {\oddsidemargin}{0in} \setlength{\textwidth}{6.2in}\setlength{\topmargin}{-0.2in}}

\def\beq{\begin{equation}\begin{aligned}}		\def\eeq{\end{aligned}\end{equation}}
\def\be{\begin{eqnarray}}  					\def\ee{\end{eqnarray}}		
   \def\bi#1{\begin{itemize}\item[#1]} 	      	   \def\ei{\end{itemize}} 
   \def\eqn#1{(\ref{#1})}
   	 \def\fn{\footnote}	\def\nm{\nonumber} 

		 \def\del{\delta}  

 \def\del{\delta}         
       \def\lam{\lambda} \def\Lam{\Lambda}

 \def\w{\omega}

 \def\HH{{\mathcal H}}      
    \def\Id{\mathbb{I}} 
 \def\pa{\partial} \def\ra{\rightarrow} 
  
 \def\dd{{\rm d}}  \def\bra{\langle}   \def\ket{\rangle}

\def\fract#1#2{{\textstyle\frac{#1}{#2}}}	 	 \def\fractje#1#2{{\scriptstyle\frac{#1}{#2}}}	
\def\ffract#1#2{\raise .2 em\hbox{$\scriptstyle#1\,$}\kern-.34 em/\kern-.34 em\lower .15 em \hbox{$\scriptstyle\,#2$}}
\def\half{\fract12}					\def\halfje{\fractje12}
\def\tl#1{\tilde{#1}} 
\def\ex#1{e^{\textstyle#1}} 			
\def\bpmatrix{\begin{pmatrix}} 			\def\epmatrix{\end{pmatrix}}
\def\bmatrix{\begin{matrix}} 			\def\ematrix{\end{matrix}} 
\def\bcenter{\begin{center}}			\def\ecenter{\end{center}}

\def\lowerheightgth#1#2#3{\(\raise-#1\hbox{\includegraphics[height=#2]{#3}}\)}
\def\lowerwidthgth#1#2#3{\(\raise-#1\hbox{\includegraphics[width=#2]{#3}}\)}
\def\widthfig#1#2{\includegraphics[width=#1]{#2}}

\def\ontt{{\mathrm{ont}}}  
  
 \def\grp{{\mathrm{group}}}

  \def\ds{\displaystyle} 
\def\weglaten#1{}	
 
   \def\ret{\\[5pt]} 

\publishversion 
 
\begin{document}
\begin{titlepage} 
\title{
 An ontological description for  relativistic,  massive bosons.
\author{Gerard 't~Hooft}}
\date{\normalsize
Faculty of Science,
Department of Physics\\
Institute for Theoretical Physics\\
Princetonplein 5,
3584 CC Utrecht \\
\underline{The Netherlands}\\
http://www.staff.science.uu.nl/\~{}hooft101 \\[15pt]
}
 \maketitle


\begin{abstract} 
Relativistic,  scalar particles  are considered, contained in a box with periodic boundary conditions. Although interactions are not expected to  be a fundamental problem, we concentrate on free particles. By considering them to be harmonic oscillators, it is found that their dynamical variables can be replaced by a completely ontological set, which means that, here, quantum mechanics does not deviate from a purely geometric, ontological particle system. The effects of the mass terms are included. Locality holds for the quantum theory, and seems to be fully obeyed also by the classical treatment, although further discussion will be needed.\\
Quantised interactions are briefly speculated on,  but mostly postponed to later. We do discuss extensively the distinction between the quantum treatment and the classical one, even though they produce exactly the same equations mathematically. We briefly explain how this result can be squared with the usual quantum no-go theorems. It is suggested to apply this theory for real time quantum model simulations. \end{abstract}

 \newsecl{Introduction}{intro.sec} 

Earlier publications\,\cite{GtHfreebosons.ref}--\cite{Wetterich2.ref} revealed how a totally classical system with a completely ontological interpretation, can be represented by a Hilbert space in such a way that it turns into a quantum theory without modifying the physical phenomena described by these equations..
 \end{titlepage} 
Here, we investigate the reverse possibility.  Do  conventional quantum systems allow for a completely classical interpretation, and under which conditions? The models proposed in Ref.\,\cite{GtHFF.ref} , require some fast moving variables on the back ground, which have to be assumed to fluctuate  thermally. Simpler models were now searched for, and found. Second quantised bosons appear to be fitting examples, since they have the right energy spectra, which automatically include  the fast variables that were required in our previous papers.

A new problem arose however: locality.\fn{The usual ``proofs" that quantum theories do not allow for a classical interpretation depend on `free will' assumptions that cannot be employed at all  in our theories.} 
 In Fourier space, the match between the classical description and a quantum formalism appears  to be perfect, but going back to position space at first sight seems to be hard. A  miracle seemed to be asked for, because the classical form would have to behave as a completely classical cellular automaton, and such a link is not easy to realise.
\setcounter{page}2
In this paper we explain how this can be done in a simple case: the \(D\)+1-dimensional system of second-quantised scalar bosons, as yet without interactions (interactions will be briefly discussed in Appendix~\ref{interact.app}). The issue encountered will be important: the classical equations of motion that we  find, at first sight deviate from locality in the sense that apparent classical interactions spread over length scales of the order of the Compton wave lengths. However, this appears to be merely the result of selecting out states with positive energy, and indeed we shall find that the squared equations  are local again, as it also happens in the conventional quantum field theories.

Our results suggest to search for promising models that may bring the Standard Model of the particle interactions in the picture. Interactions range over distance scales that are small enough to maintain the locality features of the quantum theory.

 This paper is set up as follows. 
 \def\streepje{\vskip 5pt - - - - - \vskip 5pt \noindent}
 Sections \ref{prel.sec} and \ref{setup.sec} are brief repetitions of earlier work, and may be skipped by readers who are familiar with that.
 
 But then, in section \ref{create.sec} we start our search for the ontological degrees of freedom by regarding bosons in \(D\)+1 dimensions as harmonic coupled  oscillators (taking \(D=3\) to be explicit). We explain how our search started by first focusing on the annihilation and creation operators, \(a\) and \(a^\dag\), suspecting these to have ideal commutation relations. But we spare the reader of the difficulties encountered. In an ontological theory, all observables used to describe the evolution laws (except the hamiltonian \(H\), which strictly speaking is not observable) should commute with one another, but \(a\) and \(a^\dag\) do not commute. We do see good things in \(a\) and \(a^\dag\): the way they depend on the coordinates \(q\) and momenta \(p\) does make us suspect that they do follow the circular motion of an oscillator in \(p,\, q\) space.
This was the reason to replace \(a\) and \(a^\dag\) by very similar operators \( b \) and \( b ^\dag\) that do completely commute with one another. 

 As long as we omit the interactions, a bosonic quantum field theory  is nothing but a gigantic harmonically coupled oscillator. In section \ref{massive.sec}, we search for the ontological operators for that case. Diagonalising the oscillator amounts to going to  Fourier space. This involves the momenta \(\vec k\).  At each vector \(\vec k\) we have a circular 	observable  \(\tl b \)  (the \hbox{{tilde, \(\tl{ \   }\)}}, reminds us that this field is defined in Fourier space). We Fourier transform the field \(\tl b \) back to position space, and study  its evolution.
 																																																								
 There is actually a difficulty here that we partly postpone to the discussion, section \ref{disc.sec}: the circular motion in Fourier space turns into more general motion in 3-space, which could lead to different circles when we transform back. This apparent flaw  is cured.
 
 In this section, we also ask for attention for the energy variable. In terms of the ontological observables, the energy is vaguely defined, because it does not commute with them. Therefore the question arises: how do we know when a particle is surrounded by a vacuum? As it turns out, the vacuum state is a (quantum)  distribution where all probabilities regarding the ontological variables are basically equal. 
 
 We briefly explain why our derivations are not contradicted by no-go theorems. Then, in Appendix \ref{local1.app}, we show how, in general, the evolution laws for the \( b \) fields can be transported to position space, and in Appendix \ref{local2.app} we carefully analyse what we mean by locality, and how it has to be distinguished from causality. It is important here to realise that our evolution equations are compatible with the second-order classical field equation for the classical version of the bosonic field theory.
 
 It was not our intention to discuss at length our ideas about dealing with interactions in terms of the philosophy  advocated in this paper. There are very important issues here. One might suspect that interactions will be beyond reach, but this we do not believe. There seems to be a very natural and physically appealing way to address interactions. and this idea is touched upon in  Appendix \ref {interact.app}.
 
 Fermions have not yet been considered by this author. Some first steps in this interesting field are described by Wetterich\,\cite{Wetterich.ref}\cite{Wetterich2.ref}.
 
 Our insights in this subject are still evolving, and further discussion is kept for future publications.
  
 \newsecl{Preliminaries}{prel.sec}
 
 Let us begin with recapitulating what we have found before\,\cite{GtHfreebosons.ref}--\cite{GtHCAI.ref}.
 Any elementary deterministic\fn{In some of my work I characterised  the concept of determinism to be used, as: `it's determinism all the way'. But in science fiction literature and films, authors can go way beyond that point. They assume highly conspirational modifications of ordinary logic that prevent someone to go back to the past to kill his grandfather. `Determinism all the way' stops well before such stages are reached: we accept no modification of cast-iron, down-to-earth logic.}, periodic  process can be modeled in terms of a sequence of \(N\) classical states, to be indicated in terms of Dirac kets. The evolution law can be characterised as
 	\be |0\ket\ra|1\ket\ra \cdots |N-1\ket \ra |0\ket\ra . \eel{sequence.eq}
 One may consider the continuum limit \(N\ra\infty\), which can also be written as
 	\be \frac{\dd |x\ket}{\dd t}=|f(x)\ket\ . \eel{contseq.eq}
 In the language of quantum mechanics, this reads as:
 	\be\frac{\dd \hat x}{\dd t}=-i[\hat x,\,H]\ , \qquad H=\hat p\,f(\hat x)\ , \eel{detham.eq}
where coordinates \(x\) and momenta \(p\) have been re-interpreted as being operators, indicated by a caret  (\(\,\hat{} \,\)) .  The ordering of \(\hat p\) and \(f(\hat x)\)  in Eq.~\eqn{detham.eq} may be chosen 
freely, but symmetrisation of the product would be preferred, as it would make the hamiltonian Hermitian.
Eqs.~\eqn{contseq.eq} and \eqn{detham.eq} may have non periodic solutions, while Eq.~\eqn{sequence.eq} is usually periodic. In this case, we must take our classical equations as being less-than-perfect approximations of the quantum mechanical ones. For the time being, we consider only exactly periodic processes. 

 If \(f(x)\) is known, Eq.~\eqn{detham.eq} may seem to be useful to build quantum mechanical models out of deterministic building blocks, but then there is a serious complication: quantum mechanical hamiltonians practically always have a ground state, the state of  lowest energy, which can often be interpreted as the vacuum state. In contrast, the hamiltonian \eqn{detham.eq} often has no ground state, because the spectrum of the momentum operator 
 \(\hat p\) usually is unbounded both above and below. But already in sect.~\ref{setup.sec}, we'll see a counter example.
 
 The case of finite \(N\) , Eq.~\eqn{sequence.eq}, can also be written in hamiltonian form. Here, we first define the evolution operator for one elementary time step  \(t\ra t+\del t\), a unitary operator \(U(\del t)\) defined by
 	\be |x\ket_{t+\del t}=U(\del t)|x\ket_t\ ,\eel{evolve.eq}
 where \(U(\del t) \) is an \(N\times N\) matrix having one entry 1 in each row and in each column, while all remaining entries are 0.
 
 So far, we considered only the case where there is exactly one cyclic orbit. Now if we allow for more complicated evolution laws, solutions with different periodicities may occur. These can be formally disentangled
 to yield simply different, unmixed vector spaces, without the possibility to hop from one into another; if hopping were possible, the entire set would consist of different, exactly periodic cycles, and, at first sight,  systems obtained this way will not seem to be more complex to interpret than the elementary ones.

 \newsecl{Setting up our procedures}{setup.sec}
Return to the simple, discrete, periodic, system, Eq. \eqn{sequence.eq}, with \(N\) classical states.  Let us call them 
 	\be |x\ket^{\ontt}\ ,\quad x=0,\,1,\,\cdots \,, N-1  \eel{ontstates.eq}
 (the superscript `ont' stands for `ontological').
 When acting on a state \(|x\ket^\ontt\), the evolution operator \(U(\del t)\), defined in Eq.~\eqn{evolve.eq} generates the ontological state it evolves into,  after one time step \(\del t\):
 	\be U(\del t)|x\ket^\ontt=|\,x\,+\,1\!\mod N\,\ket^\ontt\ ,\qquad U(n\,\del t)=\big(U(\del t)\big)^n\ . 	\eel{ontevolve.eq}
We can always define an \(N\) dimensional vector space with these \(N\) states as an orthonormal set as basis elements.\fn{The case where different transitions take different amounts of time \(\del t\) seems to be  physically  less meaningful; in the present models, we do not consider such possibilities (One might expect them to be relevant when gravity theories are considered).} We have 
	 \be U(N\,\del t)=U(\del t)^N=\Id\ ,  \eel{UNdelt.eq}
 from which we  deduce the eigen values of \(U(\del t)\). They are 
	 \be \ex{-iE_n}\ ,\qquad E_n=\frac{2\pi n}{N\del t}\ ,\qquad n=0,\,1,\,\cdots\,,\  N-1\,. \eel{Ueigenv.eq}
As before, \(N\) is the period of the motion, that is, the number of steps needed to return to the point of departure.
 
 The basis that diagonalises \(U\) is a set of \(N\) states \(|n\ket^E\), where \(E\) stands for `energy'.
 Eq.~\eqn{UNdelt.eq} allows us to compute a hamiltonian. It is the operator with 
 eigenvalues \(E_n\) in the same basis:
	\be U(\del t)\equiv \ex{-iH\,\del t}\ ,\qquad H|n\ket^E=E_n|n\ket^E\ .\ee
The basis states themselves are uniform superpositions of the ontological states. 
From Eq.~\eqn{UNdelt.eq} we derive
	\be  E_n=n\w\ ,\quad \w=\fract {2\pi}{N\,\del t} \ ; \qquad
	|x\ket^\ontt&=&\fract{1}{\sqrt N}\sum_{n=0}^{N-1}\ex{-2\pi i nx/N\,}|n\ket^E\ ;\labell{Eonttrf.eq}\\
	|n\ket^E&=&\fract{1}{\sqrt N}\sum_{x=0}^{N-1}\ex{\,2\pi i nx/N\,}|x\ket^\ontt\ .  \eel{ontEtrf.eq}

Notice that the ontological states used as a starting point in Eq.~\eqn{sequence.eq}, were \emph{defined} to be orthonormal. This is not at all trivial, since substituting non-integer numbers for \(x\) would yield conflicting results. Eventually, one might wish to add intermediate states between the regular ones; this would require reformulating our equations. But this is usually unnecessary. 

Also note that the frequency \(E_0\) attached to the lowest energy state may be chosen arbitrarily;  here we choose it to be zero. It must be observed however, that the complex phases of all our states are to a large extent artefacts of our procedure; the ontological states were defined to be written as  positive, real valued amplitudes, but, at best, the link to quantum mechanics is as yet still artificial.

With Eq.~\eqn{ontEtrf.eq}, we find how \(H\) acts on \(|x\ket^\ontt\). We observe that this is a spectrum of energy levels that is strictly linear in the rank \(n\). The universal energy gaps one finds
are typical for the harmonic oscillator. They are related to the linearity found in Eq.~\eqn{detham.eq}, and discussed in section~\ref{intro.sec}, but now we do have a ground state.  As stated above, one may shift the entire set of energy eigenvalues by adding the same amount \(\del\w\) to all energy levels, without the slightest modification of the physical properties os the system.

One of our basic procedures will be that now we can turn these equations around, to notice that:
\begin{quote}If we have a system where the hamiltonian \(H\) is linear in the momenta \(p\), as in Eq.~\eqn{detham.eq}, we may interpret this as a continuous,  deterministic system of one or more variables \(x\) obeying Eq.~\eqn{contseq.eq}, and furthermore,  \ret
if we have a system where the hamiltonian \(H\) has a sequence of equidistant energy levels, 
with eigenstates \(|n\ket^E\), as in Eq.~\eqn{Ueigenv.eq}, then we take it to mean that the 
states \(|x\ket^\ontt\) as defined in Eq~\eqn{ontEtrf.eq} evolve ontologically, as in Eq.~\eqn{ontstates.eq}.
\end{quote}
Thus, a hamiltonian that separates energy levels in steps of size \(\w\), corresponds to a model where a variable \(x\) takes steps of one unit in the time span of
\(\del t=2\pi/\,N\w\).

The continuum limit is obtained as  \(N\ra\infty,\ \del t\ra 0\). We keep the length of the period, \(T=N\,\del t\) constant. The distance between the consecutive energy levels, given by Eq.~\eqn{Ueigenv.eq},  stays constant, but the number of levels \(N\) becomes infinite. The hamiltonian then stays non-negative, and looks exactly as that of an ordinary, unperturbed harmonic oscillator.\fn{We ignored here the zero point oscillations, which move the entire energy spectrum upwards by \(\half \w\), where \(\w\) is the angular frequency of the oscillator.  This is exactly one half of the separation between consecutive energy levels.}

The hamiltonian of a harmonic oscillator is 
\be H=\half \hat p^2+\half\w^2 \hat q^2=(n+\half)\w\ , \qquad n=0,\,1,\,2,\,\dots\eel{harmoscH.eq}
Here \(\hat p\) is the momentum and and \(\hat q\) the position operator of an oscillating object, and \(\w\) is its angular frequency. We have \([\hat q,\hat p]=i\). The number \(n\) counts its energy levels.\fn{We shall henceforth often ignore the carets \((\,\hat{}\,\)) when \(p\) and \(q\) are considered to be operators.} One can now adopt Eqs.~\eqn{Eonttrf.eq} and \eqn{ontEtrf.eq} to express the ontological states  \(|x\ket^\ontt\) in terms of the energy eigen states \(|n\ket^E\), and use the Hermite polynomials 
to relate the states  \(|n\ket^E\) to the eigen states of \(\hat p\) and \(\hat q\).

It is important to observe that both in the finite case, described by Eqs.~\eqn{sequence.eq}, and in the continuous case of the harmonic oscillator, where the limit \(N\ra\infty\) is taken, the ontological variable \(x\) runs in a circle, with angular velocity \(\w\), the same one that describes the frequency of the oscillator.


\newsecl{Creation and annihilation}{create.sec} 
Many of the elementary features of a harmonic oscillator, are reflected in the annihilation operator \(a\) and the creation operator \(a^\dag\). They are directly related to the hamiltonian, \(H=\w a^\dag a\), where \(\w\) is the angular frequency, and the original kinetic variables 
\be p=\sqrt{\frac {\w}2}(a+a^\dag)\quad\hbox{and}\quad q=\frac 1{i\sqrt{2\w}}(a^\dag-a)\ .\eel{aadagtopq.eq}
In the energy basis, where \(E=\w n\,,\) we have 
\def\mmod{\ \mathrm{ mod }\,}
\be\bra n-1|\,a(t)\,|n\ket=\sqrt n\,\ex{-i\w t}\ ,\qquad \bra n |\,a^\dag (t)\, |n-1\ket= \sqrt n\,\ex {i \w t}\ . \eel{aadagrot.eq} 

All other matrix elements of \(a\) and \(a^\dag\) in the energy basis are zero. This allows us to read off the  time dependence of \(a\) and \(a^\dag\) directly from  Eqs.~\eqn{aadagrot.eq}.
It is quite remarkable that these operators commute with themselves at different times,
\be{}\, [a(t), a(0)]=0\ ,\qquad [a^\dag(t),\,a^\dag(0)]=0\ , \eel{aacom.eq}
a property that they do not share with the momentum operator \(p\) or the position operator \(q\) for the oscillator.
It was our intention to use either \(a(t)\)   as an ontological variable, or  use \(a^\dag(t)\)\,, but \emph{not} both \(a\) and \(a^\dag\), because \([a,a^\dag]\ne 0\). However, a more troublesome feature of these operators is that their eigen values are fundamentally complex. 

At this point, we found more useful operators to describe ontological variables. We shall call these the \emph{truncated annihilation operator} \( b \) and its conjugate \( b ^\dag\). Their matrix elements in terms of the energy eigen states \(|n\ket\) have properties more closely reflecting the original equations \eqn{Eonttrf.eq} and \eqn{ontEtrf.eq}\,:
\be \bra n-1\mmod N\ \ | b (t)|n\ket\ =\ \ex{-i\w t} \ ,\qquad
\bra n| b ^\dag(t)|n-1\mmod N\ \ket\ =\ \ex{i\w t}\ .
\eel{bbdagrot.eq}
All other matrix elements in terms of the energy eigen states vanish. 

These operators have better rotation symmetry, 
\be \bra n-m\mmod N\ \ | b ^m(t)|n\ket\ =\ \ex{-i\w\, mt}\ ,\eel{bmdef.eq}
and, besides the equivalent of Eqs.~\eqn{aacom.eq}, they obey
\be{ [ b (t),\, b ^\dag(0)]}=0\ ,\qquad b ^{-1}= b ^\dag\ .\eel{bbcom.eq}
This is why both \( b \) and \( b ^\dag\) operators are more suitable for describing ontological variables: their eigen values are all confined to the unit circle. 

An important question is how to define the operator \( b (t)\) in any given harmonic operator and how to return from the canonical variables \(p(t)\)\,, \(q(t)\) and the hamiltonian \(H\) if we have some state in the \( b \) basis.
The answers are derived from the first part of this section.

First, if we have some quantum state described in terms of the harmonically oscillating variables \(p\) and \(q\), and we have the hamiltonian \(H\)\,, \eqn{harmoscH.eq}  (with the vacuum energy \(\half\w\) subtracted), we first define the annihilation and creation operators \(a\) and \(a^\dag\) as usual, see Eq.~\eqn{aadagtopq.eq}:
\be a^\dag,\ a \  =\frac 1{\sqrt{2\w}}p \pm i \sqrt{\halfje\w}\, q \ , \ee
after which we write\quad \(H=\w\,a^\dag a\)\,.

Then, we define the operator \( b \) by using Eq.~\eqn{bbdagrot.eq}, where the time dependence is already given. We see that the energy states \(|n=N\ket\) and \(n=0\ket\) are identified. In the continuum limit, \(N\ra\infty\), this feature will be insignificant, as in practice the hamiltonian is time-independent, 
and the state \(n=N\) has almost infinite energy. If this gives rise to any problem we have to carefully take this discontinuity into account.

We can actually express \( b \) directly in terms of \(a\) and \(a^\dag\):
\be  b =(1+a^\dag a)^{-\halfje}a\ ,\quad  b ^\dag=a^\dag (1+a^\dag  a)^{-\halfje}\ .\eel{bbdagdef.eq}
We could alter the order of the operators:
\be  b =a(a^\dag a)^{-\halfje}\ ,\quad  b ^\dag=(a^\dag a)^{-\halfje} a^\dag\ , \ee
but this would require an exceptional treatment for the \(n=0\) state, which is avoided in the Eqs.~\eqn{bbdagdef.eq}.

Conversely, we can start with he \( b \) operator. Given \( b \) and \( b ^\dag\) we can identify the operator \(H\) as being the one that sends \( b \) into circles with frequency \(\w\):
\be \fract{\dd}{\dd t} b =[ b ,\,-iH]=-i\w \, b \ ,\quad [ b ,H]= b \ .\eel{detevolv.eq}
{We could write \nm \be  H=-i\w\, b \frac\pa{\pa  b }\ , \ee 
which shows a relation with our first-order deterministic approach as described in section~\ref{prel.sec}. Actually, one may regard the present work as a more advanced justification for using quantum mechanics as a formalism for deterministic first order equations.

Now to define the operators \(a\) and \(a^\dag\) in terms of \(b\) and \(b^\dag\),  we just invert Eqs.~\eqn{bbdagdef.eq}:
\be a=(1+H)^{\halfje} b \ =\  b  H^{\halfje}\ ;\qquad a^\dag= b ^\dag(1+H)^{\halfje}\ =\ H^{\halfje} b ^\dag\ . \ee 
The operators \(p\) and \(q\) now follow from Eqs.~\eqn{aadagtopq.eq}.

As the equations of motion for \( b (t)\) and \( b ^\dag(t)\),
\be\fract\dd{\dd t} b (t)=-i\w\, b \ ,\qquad \fract\dd{\dd t} b ^\dag=i\w\, b ^\dag\,, \ee
 do not involve any non-commuting operators anymore, we herewith established a non quantum mechanical description, or interpretation, of the harmonic oscillator.

Now how can we employ this observation in quantised field theories? Aren't they harmonic oscillators as well? This question is answered in section~\ref{massive.sec}.

We end this section by calculating some matrix elements for  the operator \(\hat b\). We put the caret \((\,\hat\ \,)\) on the letter \(b\) to indicate the \emph{operator} \(b\), as opposed to its eigen value which we call \(b\). The energy eigen states are \(|\,n\,\ket\) with \(H=n\w=\w a^\dag a\,\). The \(\hat b\) eigen states with eigen value \(b\) are indicated as \(|b\ket\).

We start with Eq. \eqn{bmdef.eq}. From that, we derive, 
\be\bra n+1\mmod N\,|b\ket\ =\ \bra n|\hat b|b\ket\ =\ b\bra n|b\ket \ \ra \ \bra n|b\ket= C\,b^n\ , \ee

where \(C\) is a normalisation factor. Since \(\bra N_{\,}|b\ket=\bra 0|b\ket\)\,, we have 
\be b^N=1\ \ \ra\ \ b=\ex{-2\pi i k/N}\,,\ee
where \(k\) is an integer between 0 and \(N\). These are all \(N\) eigen states of \(\hat b\).   The relation between \(k\) and \(x\) in Eq.~\eqn{Eonttrf.eq} is:
\be k=n\,x\mmod N\ .\ee
To normalise the eigen states to \(\bra b|b\ket=1\), we choose 
\be C=1/\sqrt N\ ;\qquad \bra n|b\ket=\fract 1{\sqrt N}\ex{-2\pi i k/N}\ .\ee
The exponent gives the time dependence of these eigen states. To return to the notation of 
section \ref{setup.sec}, we write the exponent as giving the time dependence of these states, exhibiting the time steps \(\del t\)\,:
\be \ex{-i\w t}\ ;\qquad \w\,\del t=2\pi/_{\,}N\ . \ee

\newsecl{Relativistic free bosons as ontological objects}{massive.sec}

Elementary particles may be regarded as wave-like solutions of a quantised field theory. They behave very much like harmonic oscillators, particularly when we switch off the effects of interactions, which would turn the oscillators into anharmonic ones.

The question we would like to ask is: can our energy spectra be compared with those of the harmonically oscillating particle fields such as the ones occurring in the Standard Model? Do they also contain sequences of equally-spaced energy levels? If so, can we also attribute an ontological interpretation to them?   And is there a chance that we can continue along this line? Would we arrive at ontologically trivial evolution equations, and could such ideas be used to study the foundations of quantum mechanics? 

In particular, we must identify   `circular orbits' along which an oscillator moves, and find out how they are to be identified as  ontological evolving variables.

The same questions were posed by this author in 2001, see Ref.~\cite{GtHfreebosons.ref}, section 3 there; but the proper answer was not discovered. In fact it seemed to be complicated indeed. The oscillator represented by free bosons seems to be an infinite set of  strongly, harmonically coupled  oscillators. How do we handle that? Now, as is often he case in our field of science, more promising  answers  are being found by going back to the basics even further. 

 In relativistic quantum field theory, at some moment in time \(t\), one introduces a field \(\Phi(\vec x)\) and its canonical associate, \(\Pi(\vec x)\). In analogy with the canonical variables \(q\) and \(p\) of the single oscillator, the commutation rules are:
 \be [\Phi(\vec x),\,\Pi(\vec x\,')]=i\del^3(\vec x-\vec x\,')\ . \ee
 
 One writes  the hamiltonian \(H\) for freely moving particles as
	\be H=\int_V\dd^3\vec x\big(\half(\Pi^2+(\vec{\nabla}\Phi)^2+M^2\Phi^2)\big)\ , \eel{hamPhi.eq}
where \(\vec\nabla=\pa/\pa\vec x\)\,, and \(M\) will be the mass of the particles.
We consider the most elementary case of spinless, massive or massless particles, as being our protoype. 
For our convenience we imagine our bosonic field theory to be confined in a box with periodic boundary conditions, defined by the periods  \(L_x,\ L_y,\ \) and \(L_z\). The volume of the box is \(V=L_xL_yL_z\).

To stay in line with the philosophy handled in the previous sections, we first have to disentangle the various oscillators, and this happens if we diagonalise the hamiltonian \eqn{hamPhi.eq}.

This diagonalisation is realised by Fourier transforming the fields in 3-space, writing the Fourier transforms by putting a tilde on top:  \(\tl \Phi(\vec k)\) and \(\tl \Pi(\vec k)\),
	\be \Phi(\vec x\,)&\equiv &\frac 1{\sqrt V}\sum_{\vec k}\,\ex{i\vec k\cdot \vec x}\,
	\tl \Phi(\vec k\,)\ , \quad\hbox{ so that}\\[3pt]	
	 \tl \Phi(\vec k\, )&\equiv &\frac {1}{\sqrt{ V}}
	 \int_V\dd^3\vec x\,\ex{-i\vec k\cdot \vec x}\,\Phi(\vec x\,)\ .
		\ee	where \(\vec k\) only takes the values such that \(k_iL_i/2\pi\) are integers.
The Fourier transforms \(\tl \Pi(\vec k,t)\) of the fields \(\Pi(\vec x,t)\) are defined in the same way.	
	
	In terms of these Fourier transforms the hamiltonian \eqn{hamPhi.eq} reads
\be H=\sum_k\half\big(\,|\tl \Pi(\vec k)|^2+(\vec k^2+M^2)|\tl \Phi(\vec k)|^2\big).\eel{fieldham.eq}
Now \(\tl{\Pi}(\vec k)\) and \(\tl{\Phi}(\vec k\,')\) all commute except when \(\vec k=\pm\vec k\,'\), in which case \(\tl\Phi\) and \(\tl\Pi\) obey exactly the commutation rules of positions \(q\) and momenta \(p\). Clearly, we have
one harmonic oscillator with \(\w=\sqrt{\vec k^2+M^2}\), at every allowed value\fn{A short-hand notation is used: \(\tl\Phi(-\vec k)=\tl\Phi^\dag(\vec k)\)\,, in all three principal directions of \(\vec k\). One can avoid this conceptual complication by replacing the exponentials with sines and cosines of \(k_i\,x_i\), at the expense of more lengthy expressions.  }  of \(\vec k\).

Figure \ref{kspace.fig} illustrates these discrete momentum values.

\begin{figure}\begin{center}
\widthfig{220pt}{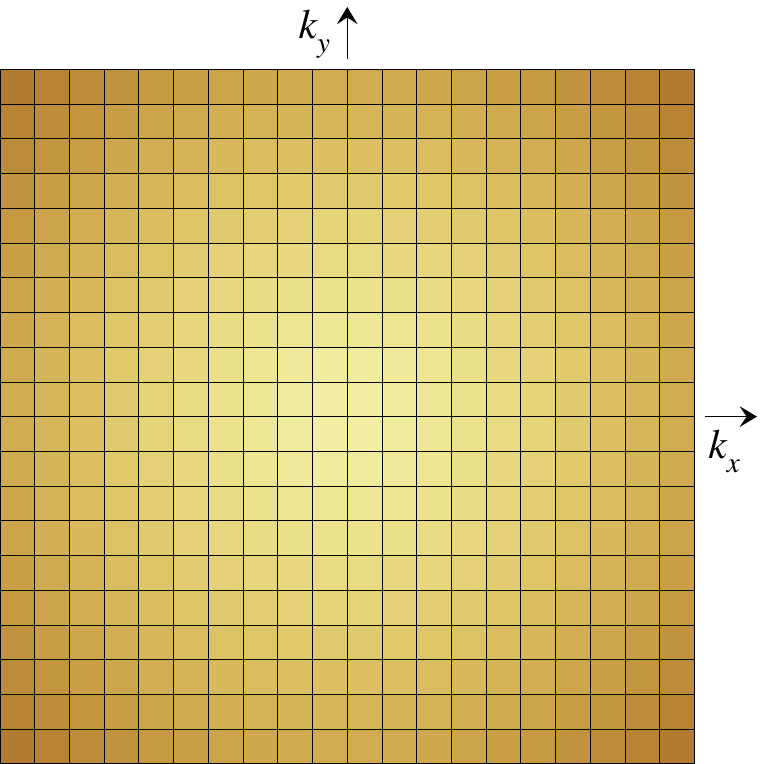}
\caption{ \labell{kspace.fig} The lattice of momentum space \(\vec k\). The size of the pixels is inversely proportional to the size of the box. This momentum space ranges to infinity in all directions. Colouring indicates the relevant time scale: \(\w =\sqrt{\vec k\,^2+M^2}\)
}\end{center}\end{figure}

Now in an earlier section, sec.~\ref{setup.sec},  we already noted  that the harmonic oscillator with frequency \(\w\) may be regarded as a classical particle rotating over a circle with angular velocity \(\w\). It is the same circular motion as what we derived in section~\ref{create.sec}. Thus, at every value of \(\vec k\), we have such an object. But what kind of classical `particle' are we looking at? How do its properties depend on the sizes \(L_x,\,L_y\) and \(L_z\)  of the box?

When \(\vec k=0\) and \(M>0\)\,, we know the answer: a `particle' runs along a circle. But this circle appears to be localised in some interior space.  Its angular velocity \(\w\) is \(M\). It appeared to be difficult to understand how it is localised in physical space-time, if really all we have is these oscillators in Fourier space.

The position of the `particle' on the circle is a physical parameter. We have one such parameter for every allowed value of  \(\vec k\), but now comes a difficulty: at   \(\vec k\ne 0\)\,, the angular frequency increases:
	\be \textstyle \w =\sqrt{M^2+\vec k^2}\ . \eel{angvel.eq}
What kind of ontological orbit gives this frequency? During this investigation, our first attempt was to find orbits of particles along some closed paths, inside our periodic box, but this did not produce physically sensible scenarios. A much more successful interpretation involves the introduction of an abstract, but entirely ontological field \(\tl  b (\vec k)\) (We put a \hbox{tilde, \(\tl{\,}\),} on top of \(b\) or \(\hat b\) to indicate that this operator is defined in Fourier space).
Here, \( \tl b \) is exactly the truncated annihilation operator studied in the previous section, \ref{create.sec}. We found that this operator commutes with itself as well as its conjugate,  at other instants in time. Furthermore,  of course it also commutes with the same operator at other values of \(\vec k\). 

This is a novelty for a quantum theory, and we decided to study it from close. First of all, we can Fourier transform \(\tl b (\vec k)\) `back' to the 3 dimensional physical  space:
	\be  b (\vec x)=
	\frac 1{\sqrt V}\sum_{\vec k} \ex{i\vec k\cdot \vec x}\,\tl  b (\vec k)\, ,\eel{bFourier.eq}
We  also introduce the truncated creation  operator \({\tl  b }^\dag (\vec k)\) , and Fourier transform that to position space,
	\be  b ^\dag(\vec x)=
	\frac 1{\sqrt V}\sum_{\vec k}\,\ex{-i\vec k\cdot \vec x}\,{\tl  b }^\dag(\vec k)\, .
	\eel{bdagFourier.eq}
Then one may verify that all these operators commute, for instance:
\be	[\tl  b (\vec k),\,{\tl  b }^\dag({\vec k}')]=0\ ,\quad\hbox{and}\quad
[  b (\vec x),\,  b ^\dag(\vec  x\,')]=0 , \ee

Using Eqs.~\eqn{bbdagrot.eq} and \eqn{bbcom.eq} we find\fn{In Eqs.~\eqn{bbdagrot.eq} and \eqn{bbcom.eq}, we still have the matrix elements; they all share the same time dependence.} time dependence of \(\tl  b (\vec k)\):
\be \frac{\dd}{\dd t} \tl  b (\vec k,t)=-i\w(\vec k)\,\tl b (\vec k,t)\ &,& \qquad \tl  b (\vec k,\,t)=\ex{-i\w(\vec k) t}\tl b (\vec k,\,0)\ ;\labell{Fevolve.eq}
\\[3pt] {[\tl  b (\vec k),\,\tl  b (\vec k\,')]}&=&0\,\quad\forall\quad\vec k,\,\vec k\,'\ .\eel{ Fcom.eq}
This just confirms that the evolution law at any value of \(\vec k\) does not affect the laws at any other \(\vec k\,'\).
However, for the \(\vec x\) dependence the situation is a bit more complicated, since \(\w\) depends on  \(\vec k\).

In the initial phases of this research the author thought that the original operators \(a\) and \(a^\dag\) had to be used at this point, but he fact that \([a,\,a^\dag]=1\) leads to expressions that are not classical, or ontological. By switching to the truncated variables \( b \) and \( b ^\dag\) this problem could be overcome. \( b \) and \( b ^\dag\) commute.\fn{A disadvantage is that \(b\) cannot be used to define the energy ground state, while \(a\) can: \(a|0\ket^E=0\).}

But now what are the equations in \(\vec x\) space? They can be read off from the laws of Fourier transformations, starting with Eqs.~\eqn{Fevolve.eq}:
\be    \fract\dd{\dd t} b (\vec x,t)&=&  {-i} \int\dd^3 \vec y\, F_1(\vec x-\vec y)  b (\vec y,t)\ ;\labell{bevolve.eq}\\
 b (\vec x,\,t)&=& \int\dd^3\vec y\,F_2(\vec x-\vec y,t) b (\vec y,0) \,,\hbox{\qquad where}\\[3pt]
  F_1(\vec z)&=&\frac 1{(2\pi)^3}\int\dd^3\vec k\,\w(\vec k)\ex{i\vec k\cdot\vec z}\ ; \qquad \w(\vec k)\ =\ 
 \sqrt{\vec k^2+M^2}\ , \labell{F1def.eq} \\[3pt]
F_2(\vec z,t)&=& \frac 1{(2\pi)^3}\int\dd^3\vec k\ \ex{\,i\vec k\cdot \vec z  -i\w(\vec k) t} \ .
 \eel{F2def.eq}

These equations are very similar to the quantum equations, but for one fact: \emph{all fields \( b (\vec x,\,t)\) in these equations commute. }They are classical equations!

In principle, they are cellular automaton equations, but then one must note that the equations are not strictly local:
each cell `interacts' with each other cell, but the interaction strength  \(F_1(\vec z)\) decreases fast when the entry \(\vec z\ =\ \vec x-\vec y\) increases. 

In Appendix \ref{local1.app} we discuss the solution of Eqs.~\eqn{bevolve.eq}--\eqn{F2def.eq}.
In particular, the direct interaction, \eqn{bevolve.eq}, decreases exponentially as soon as  \(|\vec z|>1/M\)\,, the Compton wavelength. Therefore, the system becomes more typically classical, comparable to a cellular automaton, when we are dealing with fundamental particles with the largest values for their intrinsic mass \(M\). 

We do not have to consider only that  limit, however.
In spite of this apparent non locality of the primary equation \eqn{bevolve.eq}, one can prove that the evolution law can also be cast in a local form, by squaring the evolution equation \eqn{F2def.eq}. This results in
\be  \big(\vec\pa_x\!^2-\pa_t^2\big) b =M^2 b \ , \eel{secondorder.eq}
which is of course local. A few more remarks on these subtle features are added in Appendix \ref{local2.app}, but more discussions will be inevitable.

The most important result described in this section is that we have represented the solutions of he quantum system of free bosons in a non-quantum mechanical way, because all operators in Eqs.~\eqn{bevolve.eq}--\eqn{F2def.eq} commute with one another.

\newsecl{Interpretation and discussion of the result}{disc.sec}
	We are making a claim here that, at first sight, seems to be irreconcilable with well-known facts about quantum mechanics. First of all there is the well-known Bell theorem, \cite{Bell.ref}--\cite{Bell1987.ref} the CHSH inequality\cite{CSHS.ref}, and other remarkable `quantum' features\cite{GHZ.ref}, \cite{CK2008.ref}. 
	
	Theorems such as Bell's theorem were challenged by many researchers, such as L.~Vervoort\,\cite{Vervoort2013.ref}, who seems to have mechanisms in mind that resemble what will be described here. 
	
	The fact that determinism or some version of `superdeterminism' must be considered more carefully was recently advocated by Hossenfelder and Palmer\,\cite{HossenPalmer.ref}\cite{Hossenf2.ref}.
	
Our point will be that Bell and his followers made the basic  assumption that an observer, at his or her free will, may choose which Hilbert space  components of a physical state to observe. One then can concoct a configuration where the observed object is in an entangled state, a state that cannot be realised in classical systems. 
		
The calculations provided by conventional quantum mechanics are precise and unambiguous, but their results are always expressed in terms of probability distributions.
The apparent contradictions come about when quantum physicists do not seem to understand the classical model that they claim to pulverise. The models there are quite different from what we propose. The ontological models described in our recent papers are ones that will never produce probabilities different from zero or one, \emph{if the initial state is ontologically precisely specified.} The entangled photons in Bell's experiment and others, are defined by the complex probability distributions they generate. 

If the final state has a non-trivial probability distribution, this must be due to the fact that also the initial states are not in a single ontological mode. To describe the correct initial state is impossible in practice, because at all values of the Fourier parameter \(\vec k\) we have to describe the \emph{ontological} state of that
oscillator, while, at large enough values for \(|\vec k|\), this is impossible in practice: the oscillations there are too fast. 

Instead, we usually state how many photons there are at any momentum state \(\vec k\). Usually the answer to that is that we have zero photons at the  higher energy states.

The state with zero photons is not at all ontological; to the contrary, it is an even distribution  of \emph{all possible} ontological configurations. This is because the zero photon state is an energy eigen state, which as such smears the situation over all different times.
		
In classical physics, it is obvious that one can generate `uncertainties' by not giving the proper, complete, description of an initial state. This is a general feature of classical models that quantum physicists often do not consider.

In our paper it is shown that, by taking such complications into account, one can explain `quantum behaviour' exactly. We found a basis of Hilbert space by searching for a set of mutually commuting operators. This is not as difficult as it sounds. It is done here. In our models, one can, in principle, choose a completely certain initial state. Then the final state will be a single, certain outcome as a result.

Why can experimenters not choose such an initial state? Simply because there are far too many ontological modes to consider. If we choose a beam  of particles, it is surrounded by an ocean of ontological data  that we cannot all take into account. In practice, we put all these unseen objects in their zero energy state. That is the vacuum, a highly mixed state with probability contours that are totally flat, far from what would be needed to generate final states without uncertainties.

We borrowed from quantum theory the concept of the vacuum state. Since the ingredients in this quantum state interact with the particles that we study, one finds that the results of experiments now require auxiliary states that are non commuting. In other words, the fact that we end up using non-commuting operators is caused by our inability to control everything; if we could control every single degree of freedom, we would not need quantum mechanics.

In appendix \ref{local1.app} we also derive the speed by which this automaton emits signals (over distances much larger than the Compton wavelength, finding the speed to be exactly the group velocity of the quantum wave equation, which is slower than light.

\emph{The vacuum state is the completely uniform probabilistic distribution of \emph{all} possible realistic states.} 

Now this is the way we justify our limit \(\Lam\) for the momenta \(\vec k\) in Fourier space: at all momenta \(\vec k\) with \(|\vec k\,|>\Lam\), it is assumed that the state there is not known. All possible real states have equal (low) probabilities. When we do experiments, it will always be useful to consider these unobserved states as being in the \(n=0\) state, where the quantum wave function \({}^E\bra 0|x\ket^\ontt\) takes the same value at all values of \(x\).

Thus we see that, before trying to interpret the solutions that we constructed, only the particles actually observed, stick out as being much more probable than all other states. It is their wave functions that have to be different from the universally flat wave functions. Therefore, quantum mechanics can now be seen as describing the salient probability distributions needed for describing an experiment.

This brings us to our previous publications\,
\cite{GtHFFQM.ref}\cite{GtHFF.ref} where we came close to saying the same things. There, what we needed was some variable that carries very fast fluctuating realistic variables. By having these interact with the particles that we wish to study at much lower energies -- read: longer time scales -- one introduces the effects of genuine quantum states to the dynamical phenomena of slow particles. The slow particles will cease to behave as classical objects because they are coupled to quantum objects. The quantum objects are just the high energy quantum states of the \(|\vec k\,|>\Lam\) variables of a particle type that only needs to be a virtual particle, something comparable to a super high mass unifying gauge particle. We do not see such particles directly, but we see how their interactions generate quantum features in all other, slow, particles in our world.
 	
Returning to our cellular automaton, we see that we did not copy the discrete nature of the usual automata speculated on by other researchers\,\cite{Zuse.ref}\cite{Wolfram-2002.ref}\cite{fredkin.ref}. Our present evolution law is as classical as a cellular automaton, but it is not discrete. Making the step to discrete physics should not be too difficult conceptually. We may suspect that, at the Planck scale, all physics might be discrete. Then, our construction will automatically lead to a `real' cellular automaton.

The thing to do at the point we are now, is to study other particles, notably fermions\,\cite{Wetterich2.ref}, and to study interactions. Of course, models adhered to presently , such as the Standard Model of the fundamental particles, have interaction terms. It is well known how to describe these, and how to measure them. Conversely, we may study classical interaction phenomena that will turn our system of bosons into interacting bosons. The result of these considerations will not automatically lead to a complete one-to-one comparison between classical and quantum field equations, but it seems fair to suggest to make attempts. This author is optimistic about the future of these attempts. We continue on the topic of interactions in appendix \ref{interact.app}.

\appendix
\newsecl{The classical interaction \(F_i(\vec z\,)\)}{local1.app}

In order to understand the physical features of our calculation, we must calculate the functions \(F_1\) and \(F_2\) in Eqs.~\eqn{F1def.eq} and \eqn{F2def.eq}. The equations are both of the form
\be F(\vec x)&=&\int_{|k|<\Lam}\dd^3 \vec k\ f(\vec k\,^2)\ \ex{i\vec k\cdot \vec x}\ . \ee 
 In view of rotation symmetry, we may assume \(\vec x\) to point in the 3-direction, and   the component of \(\vec k\) parallel to \(\vec x\) is called \(k\). Orthogonally to \(\vec x\) is a 2-vector \(\tl\ell\)\,:
\be \vec x=\bpmatrix \tl 0\\z\epmatrix\ ,\qquad \vec k=\bpmatrix \tl\ell\,\\ k\epmatrix\ .\ee
Introducing a temporary cut-off \(\Lam\), and writing \(|\tl \ell|\, =\ell\,\), we have
\be F(\vec x)&=&\int_{|k|<\Lam}\dd k\int_0^{\Lam^2-k^2}\dd \ell\ 2\pi\ell\,
f(k^2+\ell^2)\,\ex{ikz} \ . \\  
 F(\vec x)&=&\pi\int_{|k|<\Lam}\dd k\int_0^{\Lam^2-k^2}\dd y\,
f(k^2+y)\,\ex{ikz} \ = \ \pi\int_{|k|<\Lam}\dd k\ 
g(y)\Big|_{y=k^2}^{\Lam^2}\,\ex{ikz}\,. \eel{gkint.eq}
where\( \quad \ell\,^2=y\,;\qquad  f(y)=\ds \frac{\dd g(y)}{\dd y}\)\ .

The computation of the case \(F(\vec x)=F_1\) is arrived at by choosing 
\be f(k)=\w(k)=(k^2+M^2)^{1/2}\,,
\quad g(y)=\fract 23(k^2+M^2)^{3/2}\,, \ee 
so that
\be F(\vec x)=\fract{2\pi}{3}\int_{k=-\Lam}^\Lam\dd k(y+M^2)^{3/2}\Big|_{y=k^2}^{y=\Lam^2}\ex{ikz}\ .\ee

Assuming \(z>0\), we can deform the integration contour to run from \(k=i\Lam\),  around the
pole at \(k=iM\), back to \(i\Lam\) to get two integrals from \(k=iM\) to \(i\Lam\)\,. Writing \(k=ip\),
we then get\, (the \(\Lam\)- dependence quickly disappears):
\be (2\pi)^3
F_1(\vec x)&=&\frac{2\pi}{3}\int_M^{ \infty}i\dd p\,
\ex{-pz}\,\big\{(\Lam^2+M^2)^{3/2}-(M^2-p^2)^{3/2}\big\}\ =\labell{feerste.eq}\\
&=&-\frac{2\pi}{3}\int_M^{\infty}i\dd p\,   
\ex{-pz}\, (p^2-M^2)^{3/2}\ \big( \ex{\fract 32 i\pi } -\ex{- \fract3 2  i\pi}\big)\ =\\
&=&-\frac{4\pi}{3}\,\int_M^\infty\dd p\,\ex{-pz}\,(p^2-M^2)^{3/2}\qquad
\hbox{where } \quad z=|\vec x\,|\ . \eel{F1kernel.eq}

The function \(F_2(\vec x\,)\) is obtained in a similar way:


Here, the function \(f(\vec k)\) is 
\be \ex{-i\w(k)t}&=&\ex{-it\sqrt{k^2+M^2}}\ ;\qquad y=k^2\ , \\ 
g(y)&=&-\frac 2{t^2}\,\ex{-it\sqrt{y+M^2}}(-it\sqrt{y+M^2}-1)\ ;\eel{g2def.eq}

The integral \eqn{gkint.eq} shrinks rapidly except when the fluctuations in the \(z\) variable cancel the fluctuations in the \(t\) variable. In Eq.~\eqn{F2def.eq} and \eqn{g2def.eq}, consider a small region \(\del \vec k\) in \(\vec k\) space contributing to the integral \eqn{F2def.eq}. We look at one point \(z\) in 3-space and one moment \(t\) in the elapsed time.  The exponent then approaches the integral of
\be \ex{i( \vec z\cdot \del \vec k-t\,\del\w)} \ee
This quantity balances to a stationary point if
\be \frac{\dd \w}{\dd \vec k}=\frac{\vec z}{t}\ ,\ee
This means that points where the integral becomes substantially enhanced occur where 
\( {\vec z}/{t}={\dd\w}/{\dd \vec k}\,, \)\, or:\ret
\quad\emph{The signal expands with velocity equal to what is well-known as being the group velocity;}
for relativistic bosons, this velocity is
\be \vec v\,^\grp=\frac {\vec k}{\sqrt{k^2+M^2}}\,,\ee
which is always slower than light. 

\newsecl{Locality}{local2.app}

Locality is a statement not about the data, but about the evolution law believed to explain the data. If, to compute the predicted data in some region, the only knowledge required is the data in its immediate past, the evolution law may be declared to be a local and causal one. In   Figure \ref{local.fig}, it is surmised that all data within the past light cone  shown, populating region \(A\), suffice to predict what happens in regions \(C\) and  \(B\), up to the point \(Q\). One then does not need to know the particles that enter the regions from outside, because their effects have already been accounted for.  If the boundaries are local light cones, this law may well be relativistically invariant. In many cases, the velocity by which data can be transferred is limited by the velocity of the messenger particles, which must be going slower than light. But this velocity, called group velocity (as it can be computed by considering pakets of de-Broglie waves, which tend to form groups, with their own computable velocities), often depends on the particles concerned, and as long as it does not exceed the velocity of light, those details have little effect on the locality property.

According to this definition of locality, not only the quantum theory of free bosons obeys locality,  but also the ontological variables studied in this paper, since they are directly related. These data  are also ontological. They can all be computed, first by identifying all values of \(\vec k\) being involved, all of which are represented if we consider regions larger than the inverse of the mass \(M\).

\begin{figure}[ht] 
\begin{center} \widthfig{200pt}{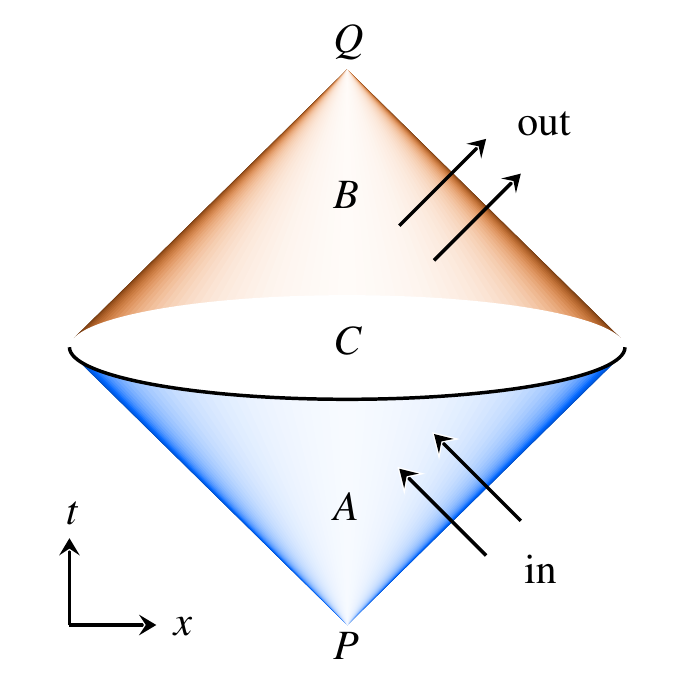}  
   \caption{\small Locality.} 
 \labell{local.fig}
See text.
 \end{center}
\end{figure}

At first sight, there might be a restriction. Our evolution equations, in particular Eq.~\eqn{bevolve.eq}, seems to transport information instantaneously, since the function \(F_1(\vec x-\vec y\,)\) connects space-like separated points, even if these points cannot be further apart than a few Compton wavelengths. However, the resulting law over   intervals  in   time-like directions, Eq.~\eqn{F2def.eq} does not require knowledge over any space-like separations, as is well-known in ordinary field theory. The proof of this is as follows. Consider the vector \(\vec k\) to be in the 3-direction. Let us shift the contour of \(k_3\) to become 
\be &&k_3=p+ia\ , \qquad p,\ a  \ \hbox{ real}\ ,\\
 && z>t\ ,\qquad \w=\sqrt{(p+ia)^2+M^2}\ = q+ib\ , \quad M,\ q,\  b\  \ \hbox{ real }; \ee
then it is easy to prove that 
	\be b<a\ . \ee

Then when we send the contour to the top of the complex plane, \(a\ra\infty\)\,, the integration over \(k_3\) converges to zero. This is because, as long as \(z>t\), the imaginary part of the exponent 
\(ik_3 \,z-i\w(k) t\) keeps the same sign and tends to infinity.

Indeed, being continuous rather than discrete, gives us this important advantage: automatic Lorentz invariance of the entire evolution equation over    finite distances in space-time.

\newsecl{Interactions}{interact.app}
The field \(b(\vec x,t)\) indeed obeys a local equation, \eqn{secondorder.eq}. This equation happens to coincide with the classical equation that was at the source of the definition of our quantum field theory. 
And indeed, by replacing this equation with the non-local, first order equation, Eqs.~\eqn{bevolve.eq} and \eqn{F1def.eq}, one can select a special class of solutions that obey positivity. 
The time derivative now has a fixed sign, so that the emerging particles are guaranteed to have positive energies. This is an encouraging sign that these equations combined generate the correct ontological solutions. 
Since all operators \(b(\vec x,t)\) commute with one another, we have a \emph{classical} theory here.

\def\free{{\mathrm{free}}}

At first sight, to introduce interactions at this level might seem to be too audacious, but there is hope. 
Our general philosophy is simple: \begin{quote} If the classical free field theory generates quantum effects simply by choosing the `chaotic' boundary conditions at high \(\vec k\) values, 
then why should the same not also be true if we add the usual interaction term to Eq.~\eqn{secondorder.eq}?\end{quote}
\be (\vec\pa_x^2-\pa_t^2)b(x)=M^2b(x)+\fract 1{6}\lam\, b^3(x)\ ,\qquad x\equiv (\vec x,t)\ . \eel{anharmonic.eq}
This is undoubtedly a totally classical equation, but one may suspect that  we can transform it
back to the quantum model, first only in the unperturbed sector by writing
\be b(\vec k)=G\,a\ ,\qquad b^\dag(\vec k)=a^\dag \,G\ ,\qquad G=
\big(1+\HH^\free(\vec k)\big)^{-\halfje}\ . \ee
The unperturbed theory has
\be \fract{\pa}{\pa t}b(\vec k)= -i\w(\vec k)b(\vec k)
\ee
To compare the effects of an interaction with the corresponding quantum theory, we must use the same strategy in both. Here this means: perform the perturbation expansion with respect to the small parameter \(\lam\) in Eq.~\eqn{anharmonic.eq}. The equation is of course classical, but this is what we are looking for. If the effects of \(\lam\) shrink when we go to high energies, we may use the same boundary conditions at high energies as in quantum field theory. Therefore we expect our mathematical method to work particularly well in asymptotically free theories such as Quantum Chromodynamics, QCD.

It is conceivable that our ideas may lead to new alleys towards efficient lattice simulation methods for such theories. There would be one great advantage: in quantum theories our refuge into perturbation expansions is almost inevitable. Perturbation expansions, however, are fundamentally divergent when applied to quantum field theories. In the ontic theories we can also do the perturbation expansion, but here, we do not have to, and we can set up simulations in  lattice configurations, using the classical equation \eqn{anharmonic.eq} non-perturbatively. There is no danger for divergences in the underlying classical models.

More speculative is the notion that the necessary discretisation of classical equations such as Eq.~\eqn{anharmonic.eq} may also lead to discretisation of   perturbation parameters such as \(\lam\), and other expansion parameters in the Standard Model.

 \end{document}